\documentclass[prd,twocolumn,nobibnotes,nofootinbib,a4paper,showpacs,showkeys]{revtex4}
\usepackage{amsmath,amssymb}
\usepackage{epsfig}
\usepackage{graphicx}

\newcommand{\pslash}{\not{\hbox{\kern-2.pt $p$}}}
\newcommand{\kslash}{\not{\hbox{\kern-1.5pt $k$}}}
\newcommand{\qslash}{\not{\hbox{\kern-1.5pt $q$}}}
\newcommand{\lslash}{\not{\hbox{\kern-.3pt $l$}}}
\newcommand{\llslash}{\not{\hbox{\kern-.3pt $l_{1}$}}}
\newcommand{\lllslash}{\not{\hbox{\kern-.3pt $l_{2}$}}}

\begin{document}
\title{\bf Diffractive Higgs boson photoproduction in peripheral collisions}
\author{G. G. Silveira}
\affiliation{Instituto de F\'{\i}sica, Universidade Federal do Rio Grande do Sul, Caixa Postal 15051, 91501-970 - Porto Alegre, RS, Brazil.}

\author{M. B. Gay Ducati}
\affiliation{Instituto de F\'{\i}sica, Universidade Federal do Rio Grande do Sul, Caixa Postal 15051, 91501-970 - Porto Alegre, RS, Brazil.}

\begin{abstract}
An alternative process is proposed for the diffractive Higgs boson production in peripheral $pp$ collisions, exploring it through the photon-proton interaction by Double Pomeron Exchange. It is estimated the event rate of the diffractive Higgs production in central rapidity for Tevatron and LHC energies, being of the order of 1 fb, in agreement to the predictions from other diffractive processes. The results are confronted with those obtained from a similar approach of the Durham group.
\end{abstract}



\maketitle

\section{Introduction}
The detection of the Standard Model Higgs boson will be the main goal of the LHC. The lower bound on the Higgs mass was estimated experimentally being $M_{H} \gtrsim 114.4\mbox{ GeV}$ with 95\% confidence level \cite{lep}. A large set of possible discovery channels was studied (see \cite{carena,hahn}), however the leading decay of the Higgs is expected to be observed as a $b\bar{b}$-pair in the mass range $M_{H} \lesssim 140\mbox{ GeV}$. A possible production of the Higgs boson under study is the diffractive proccess by Double Pomeron Exchange (DPE) \cite{bialas} in $pp$ collisions.

We apply the same idea of DPE interaction, similar to the Durham group, with the interaction occurring in the $t$-channel of the subprocess photon-proton instead of the proton-proton system. For this proposal, the formalism of impact factor is used to describe the splitting of the photon into a color dipole and its interaction with the proton at $t = 0$.

\section{Partonic process}\label{sec:amp}

The study of the diffractive production of the Higgs boson through the $\gamma^{*}q$ is based on the kinematic variables used in the description of the Deeply Virtual Compton Scattering (DVCS), where the splitting photon interacts with the proton exchanging a gluon ladder
\cite{frankfurt}.
The main feature present in this model of Higgs boson photoproduction is the interaction between the particles by DPE, providing the leading production vertex in the mass range where expect to observe experimentally the Higgs boson. This kind of process is more studied in peripheral collisions, where the impact parameter is larger than the sum of the radius of the colliding particles (protons).

\begin{figure}[h!]
\includegraphics[scale=00.60]{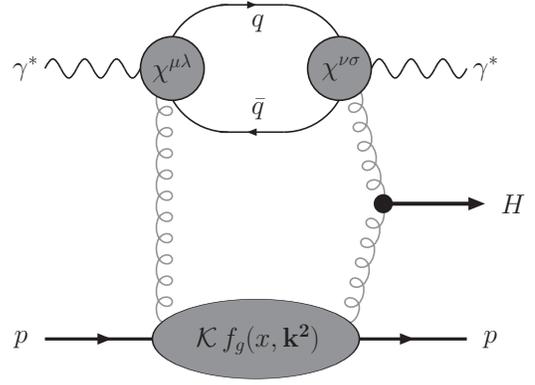}
\caption{\label{fig:foto-part} Feynman diagram representing the partonic process for diffractive Higgs production.}
\end{figure}

The Fig.\ref{fig:foto-part} shows the Feynman diagram for the subprocess of diffractive photoproduction. It represents four possibilities for the process $\gamma^{*} q$, all of them are obtained from the different coupling possibilities of the gluons to the fermions lines of the dipole.

The two-upper bubbles represent the effective vertices of the photon-gluon coupling which can be obtained through the impact factor formalism \cite{forshaw-evanson}. The same formalism is used to explore the process with a non-zero momentum transfer with two gluons exchanged in the $t$-channel. The other bubble represents the gluon density into the proton.

\begin{figure}[h]
\includegraphics[scale=00.55]{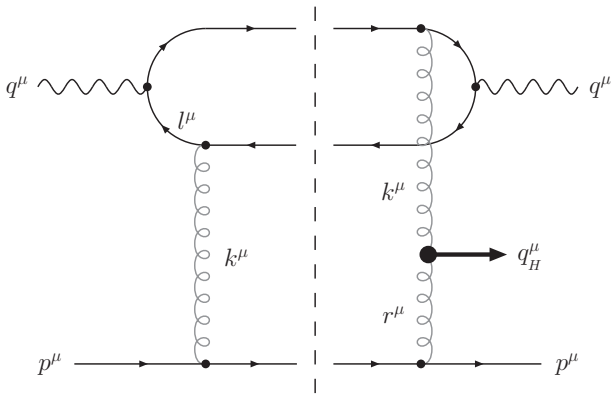}
\caption{\label{fig:foto-h} Photoproduction subprocess of the Higgs boson.}
\end{figure}

The case studied here is based on the partonic subprocess $\gamma^{*} q \to \gamma^{*} + H + q$ illustrated in Fig.\ref{fig:foto-h}. The central line cuts the diagram which represents the use of the Cutkosky rules to obtain the imaginary part of the scattering amplitude of the process through the expression
\begin{eqnarray}
{\rm{Im}}A = \frac{1}{2} \int d(PS)_{3} \, {\cal{A}}_{L} \, {\cal{A}}_{R}
\label{cut-rule}
\end{eqnarray}
with ${\cal{A}}_{L}$ and ${\cal{A}}_{R}$ being the amplitudes on the left and right side of the cut, respectively, and $d(PS)_{3}$ is the volume element of the three-body phase space. The scattering amplitude of the process is treated essentially as an imaginary quantity since the particle exchanged has the vacuum quantum numbers \cite{foldy}, so we can neglet the real one.

In the Dipole Model \cite{mueller1} the splitting of the photon into a quark-antiquark pair is treated by an wave function, where the product of this quantity with its complex conjugate represents the presence of the fermion loop.

Calculating the amplitude assigned in (\ref{cut-rule}), the product of the left and right amplitudes results
\begin{widetext}
\begin{eqnarray}
{\cal{A}}_{L} {\cal{A}}_{R} = (4 \pi)^{3} \alpha^{2}_{s}\alpha \left( \sum_{q} e^{2}_{q} \right) \left( \frac{\epsilon_{\mu}\epsilon^{*}_{\nu}}{k^{6}} \right) \frac{V^{ba}_{\delta\sigma}}{N_{c}} \left( t^{b}t^{a} \right) \, 2 \left[ \frac{T^{\mu\lambda\sigma\nu}}{l^{4}} + \frac{T^{\lambda\mu\sigma\nu}}{l^{2}(k + l + q)^{2}} \; \right] \; 4 p_{\lambda}p^{\delta}
\label{amp-cut-fhiggs}
\end{eqnarray}
\end{widetext}
where $\epsilon_{\mu}$ and $\epsilon^{*}_{\nu}$ are the polarization vectors of the initial and final photons, respectively, and $T^{ijkl}$ are the traces from the dipole. The vector $l^{\mu}$ is the four-momentum of the quark circulating in the fermion loop and $p^{\mu}$ is the four-momentum of the colliding proton. There are four different diagrams to represent the dipole in this process since the two gluons couple to its fermion lines. However, if both gluons couple to the upper fermion line, it contributes equally as the couplings to the lower one. This equivalence also occur in the coupling to distinct fermion lines. Thus, we need to take into account only one diagram with the gluons coupled to the same fermion line and other that they couple to different ones, and add a factor of 2 to include the other contributions.

The quantity $V^{ba}_{\delta\sigma}$ represents the production vertex $gg \to H$ which is known as \cite{kniehl}
\begin{eqnarray}
V^{ab}_{\mu \nu} = \delta^{ab} \left( g_{\mu \nu} - \frac{k_{2\mu} k_{1\nu}}{k_{1} \cdot k_{2}} \right) V,
\end{eqnarray}
where
$V \approx M^{2}_{H} \alpha_{s} / 6 \pi v$
being valid for the production of a non-heavy Higgs boson ($M_{H} \lesssim 200\textrm{ GeV}$).

However, the value of traces involving a product of Dirac $\gamma^{\mu}$-matrices is obtained adopting an adjusted parametrization for the four-momenta present in the process. We adopt the Sudakov parametrization where the four-momenta are decomposed under three base-vectors: two vectors of light-type $p^{\mu}$ and $q^{\prime \mu}$, with $q^{\prime \mu} = q^{\mu} + x p^{\mu}$, and a third vector lying in a plane perpendicular to the incident axis. The main kinematic variables of interest are the center-of-mass energy $s = (q + p)^{2}$ and the momentum fraction $x = Q^{2} / \, 2 \, (p \cdot q)$, with $Q^{2}$ being the photon virtuality. The decomposition permits to write the four-momenta in the form
\begin{subequations}
\begin{eqnarray}
l^{\mu} &=& \alpha_{\ell} q^{\prime \mu} + \beta_{\ell} p^{\mu} + l^{\mu}_{\perp} \\
k^{\mu} &=& \alpha_{k}q^{\prime \mu} + \beta_{k} p^{\mu} + k^{\mu}_{\perp} % \\
\end{eqnarray}
\end{subequations}

The polarization vectors do not depend on the $t$-variable, its sum being over transversal and longitudinal components expressed by the relations %\cite{forshaw-book}
\begin{subequations}
\begin{eqnarray}
\epsilon^{L}_{\mu} \, \epsilon^{L *}_{\nu} &=& \frac{4Q^{2}}{s} \, \frac{p_{\mu}p_{\nu}}{s} \\
\sum \epsilon^{T}_{\mu} \, \epsilon^{T *}_{\nu} &=& - g_{\mu\nu} + \frac{4Q^{2}}{s} \, \frac{p_{\mu}p_{\nu}}{s}.
\end{eqnarray}
\end{subequations}

In order to reach our initial proposal, it is necessary to perform the approximation on the photon virtuality taking the limit $Q^{2} \to 0$. This approximation is a realistic limit in the peripheral collisions context, in which the photon field around the hadrons is composed of real photons. Thus, we can simplify enormously obtaining the relation
\begin{eqnarray}
\label{amp-freal}
(\textrm{Im}A)_{T} = \frac{10s}{9} \left( \frac{M^{2}_{H}}{\pi v} \right) \alpha_{s}^{3} \alpha \, \sum_{q} e^{2}_{q} \left( \frac{2C_{F}}{N_{c}} \right) \int \frac{d\mathbf{k}^{2}}{\mathbf{k}^{6}}
\end{eqnarray}

We compute the cross section as a distribution in central-rapidity of the Higgs boson ($y_{H} = 0$), obtaining
\begin{eqnarray}\nonumber
\frac{d\sigma}{dy_{H}d\mathbf{p}^{2}_{H}} = \frac{25\alpha_{s}^{4} \alpha^{2}}{2^{3} \, 81\pi^{3}} \left( \frac{M^{2}_{H}}{N_{c}v} \right)^{2} \left( \sum_{q} e^{2}_{q} \right)^{2} \left[ \int \frac{\alpha_{s} \, C_{F}}{\pi} \frac{d\mathbf{k}^{2}}{\mathbf{k}^{6}} \right]^{2}
\end{eqnarray}

The main aspect obtained from this result is the sixth-order $\vec{k}$-dependence compared to the result of Durham group, which had been obtained with a fourth-order dependence. This difference appears due to the presence of the photon in the process which simplifies the result by the existence of only one parton distribution in the differential cross section.

\section{Photon-Proton Collisions}\label{sec:proton}

A realistic case of photon-proton interaction in peripheral collisions is built if we substitute the contribution of the gluon-quark vertices by a partonic distribution into the proton to express the coupling of the gluons to the proton, as illustrated by the lower blob of Fig.\ref{fig:foto-part}.

However, the condition $t = 0$ is not sufficient to determine the gluon density function and it is necessary to assume a small value to the momentum fraction in this region of interest, as $x \sim 0.01$, such that we can safely put $t = 0$ \cite{KMR-1997}.

Therefore, the follow replacement is made to describe the $\gamma p$ interaction
\begin{eqnarray}
\frac{\alpha_{s} \, C_{F}}{\pi} \;\; \longrightarrow \;\; f_{g}(x,\mathbf{k}^{2}) = K \left( \frac{\partial [xg(x,\mathbf{k}^{2})]}{\partial \ell n \, \mathbf{k}^{2}} \right)
\end{eqnarray}
where $f_{g}(x,\mathbf{k}^{2})$ is the non-diagonal gluon distribution function into the proton which evolves through BFKL equation. Assuming a non-diagonal distribution it is possible to approximate this non-diagonality by a multiplicative factor $K$ which possess a Gaussian shape \cite{shuvaevetal}
$K = (1.2)\, \textrm{exp}(-b\boldsymbol{p}_{H}^{2}/2)$, where $b = 5.5\mbox{ GeV}^{-2}$ is the impact parameter. This factor can be seen as a representation of the proton-Pomeron coupling. 

Finally, the differential cross section has the form
\begin{eqnarray}
\left. \frac{d\sigma}{dy_{_{H}}} \right|_{y_{_{H}} = 0} \sim \frac{1}{b} \left[ \int \frac{d\mathbf{k}^{2}}{\mathbf{k}^{6}} \; f_{g}(x,\mathbf{k}^{2}) \right]^{2}.
\end{eqnarray}
An important feature considered by the Durham group is the suppression of the gluon emissions from the annihilation vertex, i.e., \textit{bremsstrahlung} emissions \cite{KMR-1997}. The suppression probability for the emission of one gluon can be computed with the help of Sudakov form factors. For many emissions, this factor exponentiates and is taken into account by the introduction of an exponential factor to the gluon distribution.

A last important aspect involving diffractive processes is to compute the rapidity gaps present in the final state. These quantities are introduced in the model since the interaction between the colliding particles occurs by means of an exchange of a particle with the vacuum quantum numbers, in this case, the Pomeron. Thus we consider a multiplicative factor $S^{2}_{gap}$ to include this physical aspect. Several approaches predict this quantity \cite{KMR,gotsman}, where the survival probability for Higgs production is estimated to be 3\% for LHC ($s = 14\textrm{ TeV}$) and 5\% for Tevatron ($s = 1.96\textrm{ TeV}$).

\section{Numerical results}\label{sec:results}

Reaching our goal to build this model, we are able to calculate the differential cross section for the Higgs boson diffractive production through the $\gamma p$ interaction. Avoiding infrared divergencies we made a cut in the integration on the gluon transverse momentum \cite{forshaw-KMR}.

\begin{figure}[h]
\centering
\scalebox{00.325}{\includegraphics*[38pt,48pt][707pt,530pt]{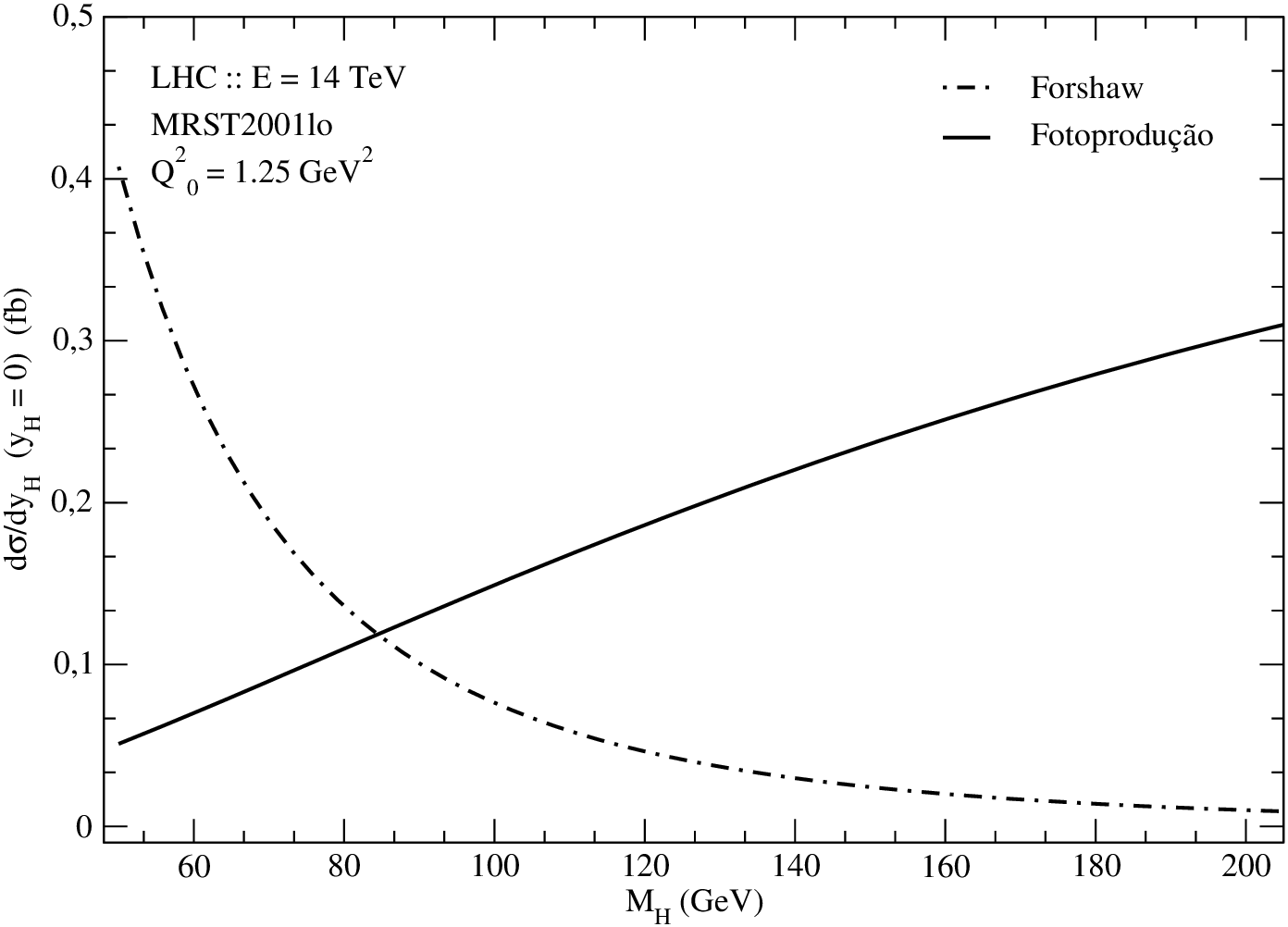}}
\caption{\label{fig:fig4} Differential cross section $d\sigma/dy_{H} (y_{H} = 0)$ for LHC energy.}
\end{figure}

The first step is to compare the results obtained with this model with those obtained before for the production in direct $pp$ collisions. For this proposal we compare our results with the results of the Durham group implemented by Forshaw \cite{forshaw-KMR}. The prediction for the differential cross section in central-rapidity for LHC is calculated using the parametrization MRST2001 in leading order approximation for the gluon distribution function taking an initial cut of $\boldsymbol{k}^{2}_{0} = 1.25\textrm{ GeV}^{2}$. The result is expressed in Fig.\ref{fig:fig4} where the differential cross section is fitted in function of the Higgs boson mass. Therefore the behavior of the photoproduction results is expected to not fit like the results of direct $pp$ collisions due to the presence of only one parton distribution in the $\gamma p$ approach.

\begin{figure}[h]
\centering
\scalebox{00.315}{\includegraphics*[20pt,48pt][707pt,530pt]{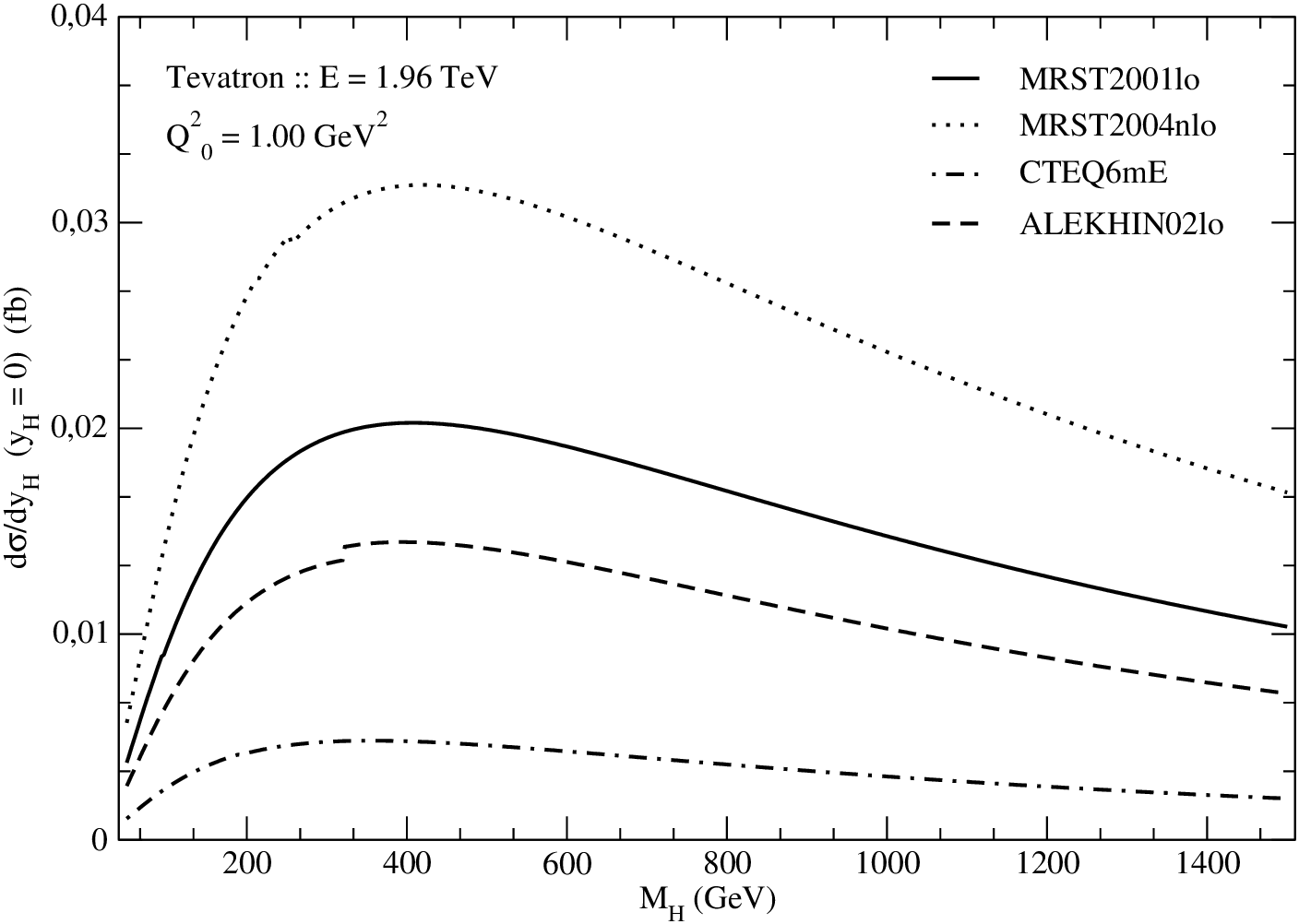}}
\scalebox{00.315}{\includegraphics*[20pt,48pt][707pt,530pt]{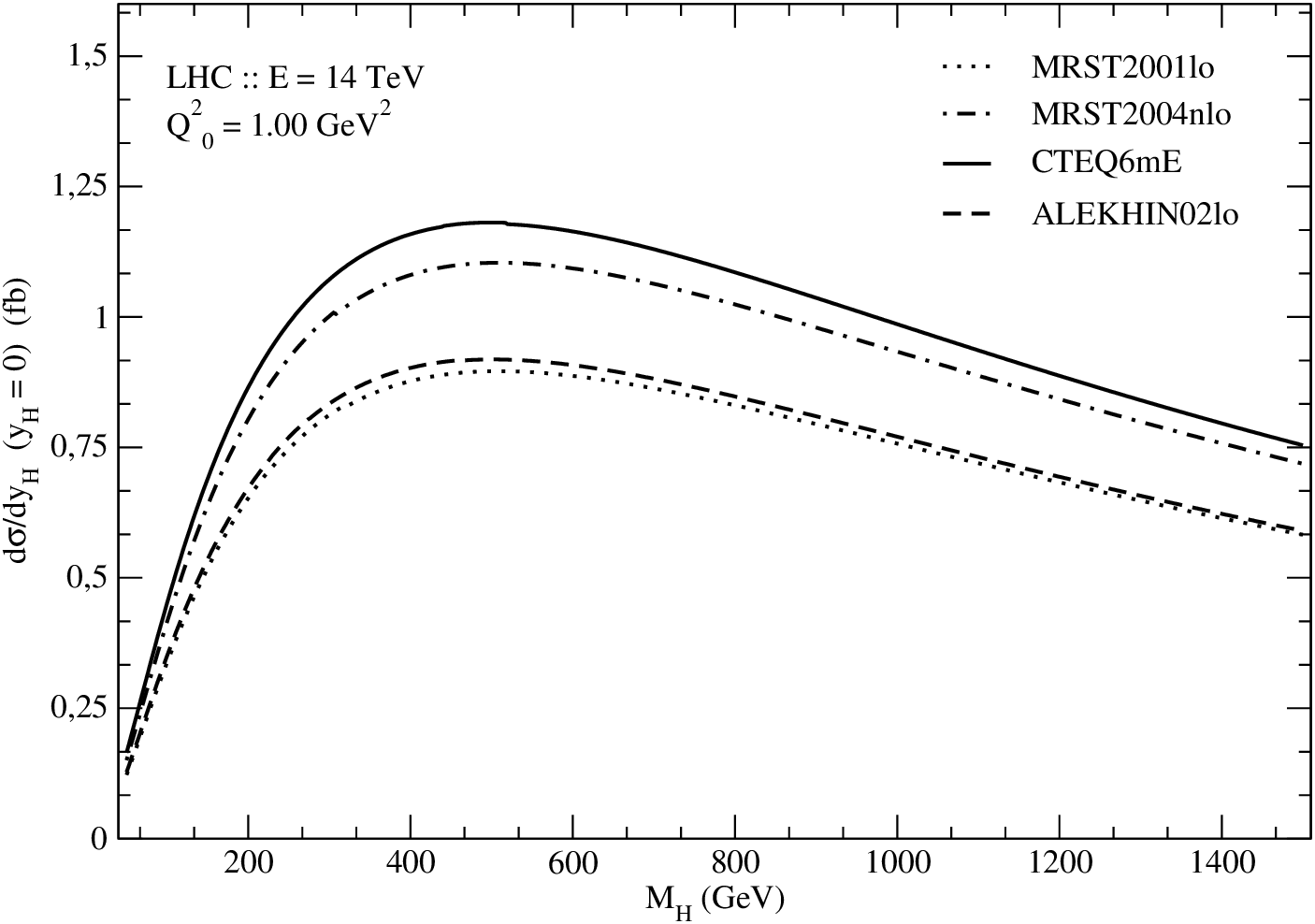}}
\caption{\label{fig:fig5} Differential cross section $d\sigma/dy_{H} (y_{H} = 0)$ for energies of Tevatron and LHC}
\end{figure}

Extending these numerical analyses we predict the differential cross section adopting some distributions functions for the gluon content into the proton, which is shown in Fig.~\ref{fig:fig5}. The non-diagonality of the distributions was approximate by a multiplicative factor which permit us to account the usual diagonal distributions. All these distributions were evolved from an initial momentum $\boldsymbol{k}^{2}_{0} = 1\textrm{ GeV}^{2}$, value adopted to be an average between the initial cuts assumed by each parametrization. As an evidence we can see a gap in the results to LO and NLO distributions in this range of energy.

\begin{figure}[t]
\scalebox{00.32}{\includegraphics*[20pt,48pt][707pt,530pt]{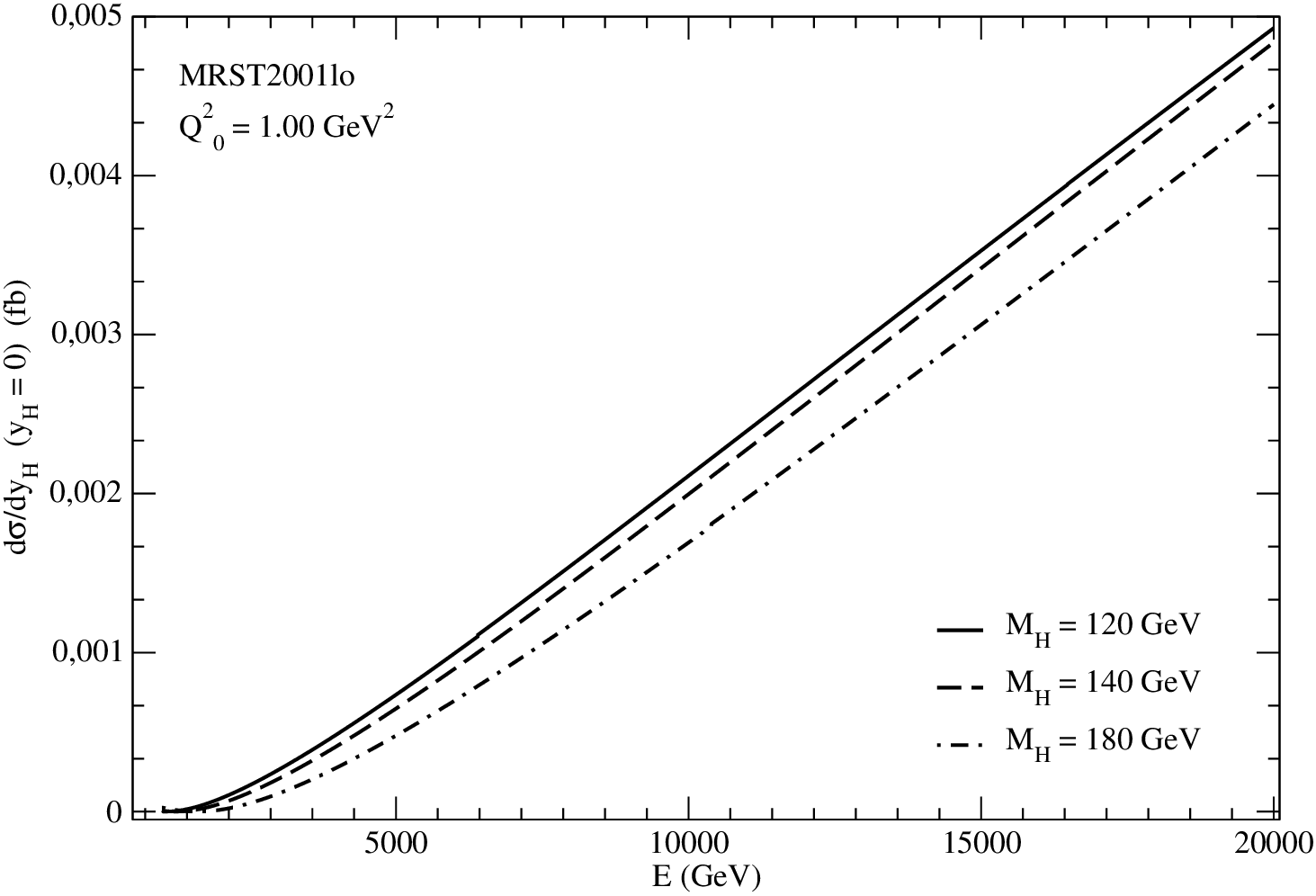}}
\caption{\label{fig6}Differential cross section $d\sigma/dy_{H} (y_{H} = 0)$ varying the Higgs mass.}
\end{figure}

\begin{figure}[t]
\scalebox{00.32}{\includegraphics*[20pt,48pt][707pt,530pt]{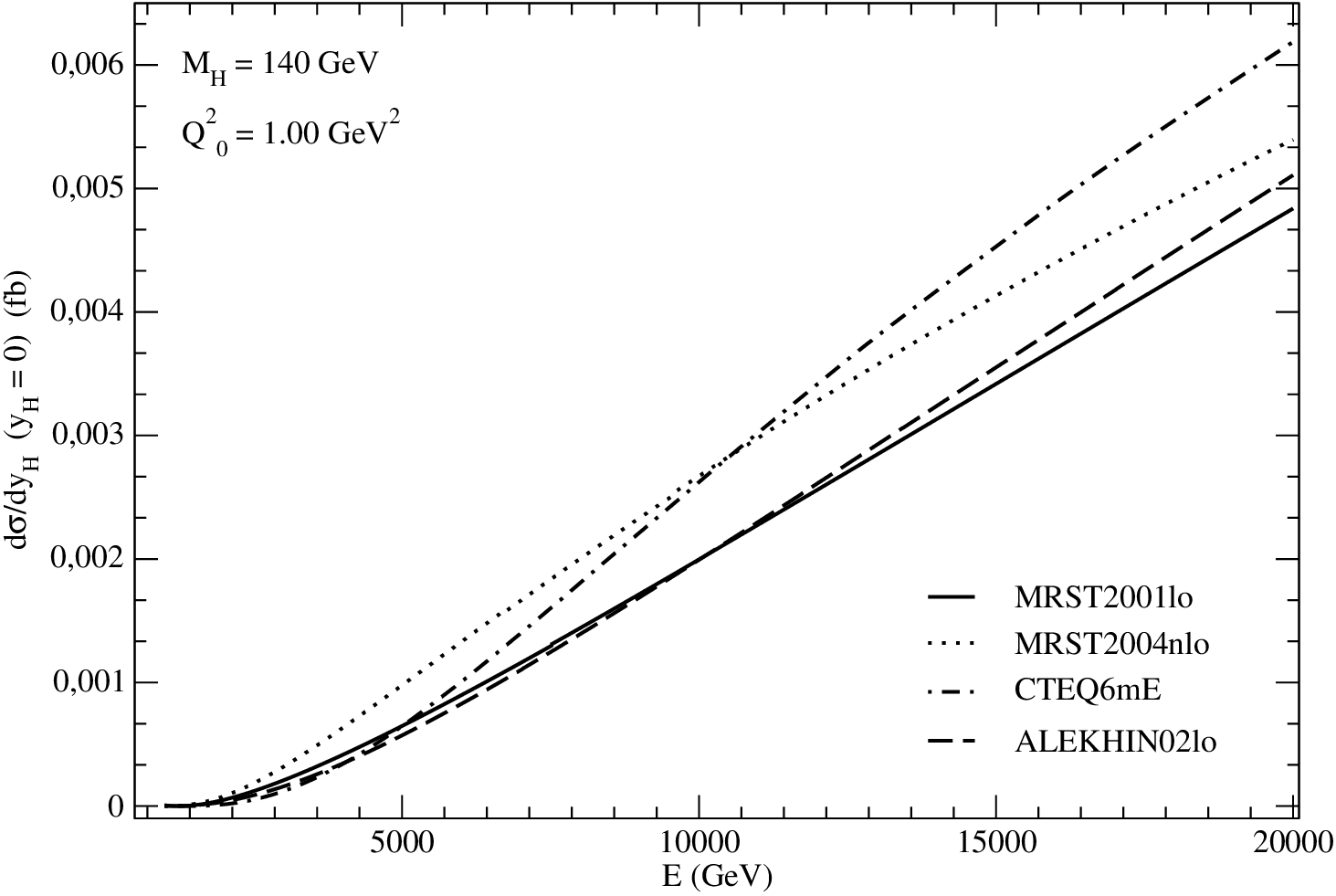}}
\caption{\label{fig7}Differential cross section $d\sigma/dy_{H} (y_{H} = 0)$ for different PDF in the process.}
\end{figure}

A final analysis consists in the observation of the dependence of the differential cross section on the center-of-mass energy inspecting it for distinct mass values of the Higgs boson and its behavior in the distribution functions. We obtain the results under three distinct values of Higgs boson mass, as shown in Fig.\ref{fig6}. The growth of the differential cross section with $s$ is linear and approximately the same for any value of Higgs boson mass. Up to energies of the order of $\sqrt{s} \approx 4\textrm{ TeV}$ we can see that the differential cross section has a parabolic shape due to its quartic dependence on the Higgs mass. The Sudakov form factors acts to increase the results stating a linear shapes for higher energies. This dependence is analysed with the distribution functions used before presenting the same linear behavior for higher energies, showed in Fig.\ref{fig7}. An important aspect observed in this second analysis is the same difference between LO and NLO distribution functions observed before.

% \vspace*{00.55cm}

\section{Conclusions}\label{sec:ccl}

The numerical results obtained in this approach demanded a more complex calculation since it was computed the dipole contribution in the Higgs boson production by DPE. As a reward we could obtain a simple result to the event rate, however from a more complex physical process than those studied by the Durham group. Performing the phenomenological analyses, the results show an event rate of the order of 1 fb, in accord to the predictions from other diffractive processes for Higgs production. We expect an agreement to the previous results of the Durham group when a more complete study be performed with the introduction of a photon distribution into the proton, and to effectively compute the Higgs production in peripheral hadron-hadron collisions.

\section{Acknowledgements}

This work is partially supported by CNPq (G.G.S. and M.B.G.D.).

\end{document}